\documentclass[a4paper,11pt]{article}
\usepackage{pos}

\usepackage{bbold}
\usepackage{physics}
\usepackage{siunitx}
\usepackage{xcolor}
\usepackage[symbol]{footmisc}

\title{Staggered fermions with taste splitting mass term on dynamical configurations}

\author[a]{Gianluca Fuwa}
\author*[a]{Nuha Georgiev-Chreim}
\author[a]{Christian Hoelbling}

\affiliation[a]{Department of Physics, University of Wuppertal, Gaußstraße 20, D-42119 Wuppertal, Germany}

\emailAdd{nuha.chreim@uni-wuppertal.de}
\emailAdd{hch@uni-wuppertal.de}

\abstract{We present numerical results of staggered fermions with a taste splitting mass term on dynamical configurations. The rise of gluonic counterterms from rotational symmetry breaking is studied for a single taste operator and the pion propagator is computed. Preliminary numerical results are given for lattice sizes up to $16^4$.}

\FullConference{The 42nd International Symposium on Lattice Field Theory (LATTICE2025)\\
2-8 November 2025\\
Tata Institute of Fundamental Research, Mumbai, India\\}

\begin{document}
\maketitle

\section{Introduction}
Dynamical fermions are common practice nowadays 
with staggered fermions still among the top preferred fermions formulations. The staggered construction \cite{Susskind,Banks} has the advantage of a small operator while preserving a proxy of the chiral symmetry. The remaining four staggered tastes can be decoupled in a similar manner as for Wilson fermions \cite{Golterman, Adams, Hoelbling,deForcrand,Durr,Misumi,Zielinski}. Following up previous results in pure gauge SU(3) \cite{PoS2024}, we present first results of staggered fermions with a taste splitting mass term on dynamical configurations. Besides the actual simulation, one of the main goals is to study the rotational symmetry breaking via gluonic operators for the single taste mass term formulation \cite{Sharpe}. Finally, we compute the pseudoscalar propagator.\\
The staggered action
\begin{align}
    S_{st} =a^4\sum_n \Bar{\chi}(n)\left(\eta_{\mu}\frac{1}{2a}\left[\chi(n + \hat{\mu}) - \chi(n-\hat{\mu})\right]  + m\chi(n))\right)
\end{align}
with spinless fields $\chi$, staggered phase $\eta=(-)^{\sum_{\nu < \mu}n_{\nu}}$, on-site mass $m$ and lattice spacing $a$ has a spin-taste basis that is given by \cite{Daniel} 
\begin{align}
    (\gamma_{S}\otimes \xi_{F})_{xy}=\frac{1}{2^{D/2}}\tr(\Gamma^{\dagger}(x)\gamma_{S}\Gamma(y)\gamma^{\dagger}_{F})
\end{align}
in leading order. The notation is such that $\xi_{F}=\gamma_{F}^{T}$, where $\gamma$ are the gamma matrices and $\Gamma(n) = \prod_{\mu}\gamma_{\mu}^{n_{\mu}}$, acts on flavor indices. From the spin-taste basis it is evident that taste splitting masses have to be spin singlet operators. General flavored mass terms, which are hermitian, are \cite{Golterman}
\begin{align}
    M= \mathbb{1} \otimes \left(m\mathbb{1}+m_{\mu}\xi_{\mu}+\frac{1}{2}m_{\mu\nu}\sigma_{\mu\nu}+m_{\mu}^{5}i\xi_\mu \xi_5 +m^5\xi_5\right).
\end{align}
and when considering only $\epsilon$-hermitian operators, one is left with \cite{Adams,Hoelbling}
\begin{alignat}{3}
    &\qquad 1\otimes \sigma_{\mu\nu} &&\longleftrightarrow M_{\mu\nu} =i\epsilon_{\mu\nu}\eta_{\mu}\eta_{\nu}C_{\mu}C_{\nu}\\
    &\qquad 1\otimes\xi_5 &&\longleftrightarrow M_A=\epsilon\eta_5C. 
\end{alignat}
in addition to the taste-singlet mass which is the usual dirac mass. The operator $M_{\mu\nu}$ is an antisymmetric 2-hop tensor with $\sigma_{\mu\nu}=i\xi_{\mu}\xi_{\nu}$, $\epsilon_{\mu\nu}=(-)^{x_{\mu}+x_{\nu}}$ and $C_{\mu}=\frac{1}{2}\left(U_{\mu}\delta_{x+\hat{\mu},y}+U_{\mu}^{\dagger}\delta_{x-\hat{\mu},y}\right)$ where $U_{\mu}$ is the usual link variable. The 4-hop Operator, also known as the Adams type, $M_{A}$ \cite{Adams} involves $\epsilon=(-)^{\sum_{\mu}x_{\mu}}$, $\eta_{5}=\eta_{1}\eta_{2}\eta_{3}\eta_{4}$ and $C=\left(C_{1}C_{2}C_{3}C_{4}\right)_{sym}$, which is symmetrized sum over all permutations of hopping parameters.\\
From the 2-hop operator, single taste mass terms can be constructed by hopping in all four directions, i.e. $M_{H}=M_{\mu\nu}+M_{\rho\sigma}$ with $\mu\neq\nu\neq\rho\neq \sigma$ permutations of $\{1,2,3,4\}$. The subscript of $M_{H}$ refers to \cite{Hoelbling} where the single taste operator was first classified and studied. 

\section{Discrete Symmetries} 
The discrete symmetries of the staggered action are given by \cite{Toolkit, LectureNotes}
\begin{itemize}
          \item
          Rotations
          \begin{align}
              R_{\mu\nu}: \chi(x) \to S_R\left(R^{-1}x\right) \chi\left(R^{-1}x\right)
          \end{align}
          \item 
          Shifts
           \begin{align}
              S_{\mu}: \chi(x) \to \zeta_{\mu}(x) \chi\left(x + \hat{\mu}\right)
          \end{align}
            \item 
          Spatial inversion
           \begin{align}
              I_S: \chi(x) \to \eta_{4}(x) \chi\left(I_S^{-1}(x)\right)
          \end{align}
            \item 
          Charge conjugation
           \begin{align}
              C_0: \begin{cases}
             \chi(x) \to \epsilon(x) \Bar{\chi}\left(x\right)\\
             \Bar{\chi}(x) \to -\epsilon(x) \chi\left(x\right)
              \end{cases}
          \end{align}
          \item 
          Taste transformation
          \begin{align}
              \Xi_{\mu}: \chi(x) \to \zeta_{\mu}(x) \chi\left(x\right)
          \end{align}
      \end{itemize}
      Discrete rotations on the lattice are by $\pi/2$ and include a factor 
      \begin{align*}
        S_R(n) = \frac{1}{2}\left(1 \mp \zeta_{\mu}\zeta_{\nu} - \eta_{\mu}\eta_{\nu}\pm \zeta_{\mu}\zeta_{\nu} \eta_{\mu}\eta_{\nu} \right)\quad (\mu \lessgtr \nu)
    \end{align*}
    Axial reversal $I_S$ reverses the $\mu$-direction 
    \begin{align*}
        x_\mu &\rightarrow -x_\mu \\
        x_\nu &\rightarrow x_\nu \quad \mu\neq\nu.
    \end{align*}
    These symmetries are partially broken when adding tasty masses. Particularly the single taste operator is of interest here as it breaks the rotational symmetry from which gluonic counterterms emerge \cite{Sharpe}. The mass term $M_{H}=M_{\mu\nu}+M_{\rho\sigma}$ is invariant under the following subgroups
    \begin{itemize}
          \item Rotation: $R_{\mu\nu}R_{\rho\sigma}$
          \item Shifts and spatial inversion: $S_{\mu} S_{\nu} S_{\rho} S_{\sigma}$ and $S_{\mu} I_S$
          \item Charge conjugation \footnote[3]{Note that when considering other single taste operators, e.g. $M_{H}=\frac{1}{\sqrt{3}}(M_{12}+M_{34}+M_{13}+M_{42}+M_{14}+M_{23})$, the charge conjugation symmetry is broken differently \cite{Misumi}.}: $R_{\mu\rho}^{2}C_{0}$
    \end{itemize}
    The breaking of rotational symmetry causes the gluonic action
    \begin{align}
        S_{G}=\frac{a^4}{2g^2}\sum_{n}\sum_{\mu\nu}\tr(F^2_{\mu\nu}) +\mathcal{O}(a^2)
    \end{align}
    to split into two parts that renormalize differently, i.e.
    \begin{align}
        a(F_{\mu\nu}^2+F_{\rho\sigma}^2)+b(F_{\mu\rho}^2+F_{\mu\sigma}^2+F_{\nu\rho}^2+F_{\nu\sigma}^2), \quad a \neq b.
    \end{align}
    These counterterms can be measured on the lattice by expressing them in terms of averaged plaquettes
    \begin{align}
        \Delta_{\mu\nu+\rho\sigma}=\frac{1}{2}\left(U_{\mu\nu}+U_{\rho\sigma}\right)-\frac{1}{4}\left(U_{\mu\rho}+U_{\nu\rho}+U_{\mu\sigma}+U_{\rho\sigma}\right).
        \label{eq:counterterm_general}
    \end{align}

\section{Numerical Setup}
\begin{figure}[h]
    \centering
    \includegraphics[width=0.8\textwidth]{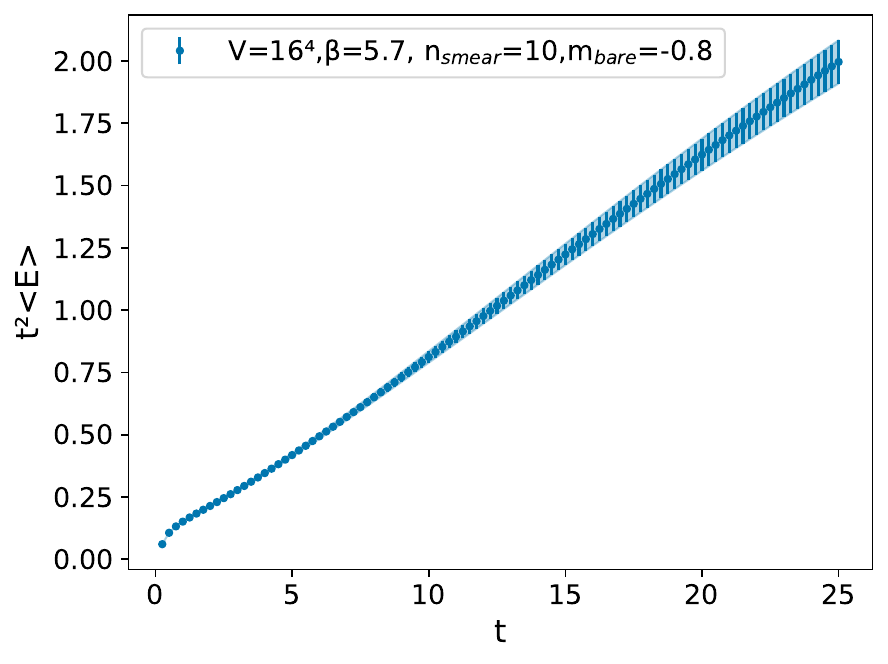}
    \caption{Scale setting via the Wilson flow.}
    \label{fig:scalesetting}
\end{figure}

The gauge configurations on which measurements are carried out are generated with the single taste operator,
\begin{align}
   D_{H} = D_{st}+M_{12}+M_{34}
   \label{eq:single_taste}
\end{align}
$D_{st}$ being the typical staggered dirac oparator, for number of flavors $N_{f}=2$ and a bare mass of $m_{bare}=-0.98$. The configurations are stout smeared with a smearing parameter of $\rho=0.12$. The lattice parameters are given in 
\autoref{tab:gitterparameter}.\\
\begin{table}[h]
    \centering
    \begin{tabular}{ c|c|c }
        \hline
        $\quad$ $L/a$ $\quad$ & $\quad$ $\beta$ $\quad$ & $\quad$ $\text{n}_{\text{smear}}$ $\quad$ \\
        \hline \hline
         $16$ & $5.7$ & $0$ \\
         & & $10$  \\ \hline 
    \end{tabular}
    \caption{Lattice parameters.}
    \label{tab:gitterparameter}
\end{table}\\
For the scale setting, we extract $t_0$ using the Wilson flow \cite{Lüscher} for $N_f=2$ and $\sqrt{t_{0,\text{phys}}}=\SI{0.1539(12)} {\femto\meter}$ \cite{Sommer}. \autoref{fig:scalesetting} shows $t^2\langle E\rangle$ vs. the flow time t where $\langle E\rangle$ is the averaged energy density for gauge configurations with $\text{n}_{\text{smear}}=10$.

The corresponding lattice spacing is $a=0.0843(29)\SI{}{\femto\meter}$.

\section{Results}
\begin{figure}[ht]
    \centering
    \includegraphics[width=0.8\textwidth]{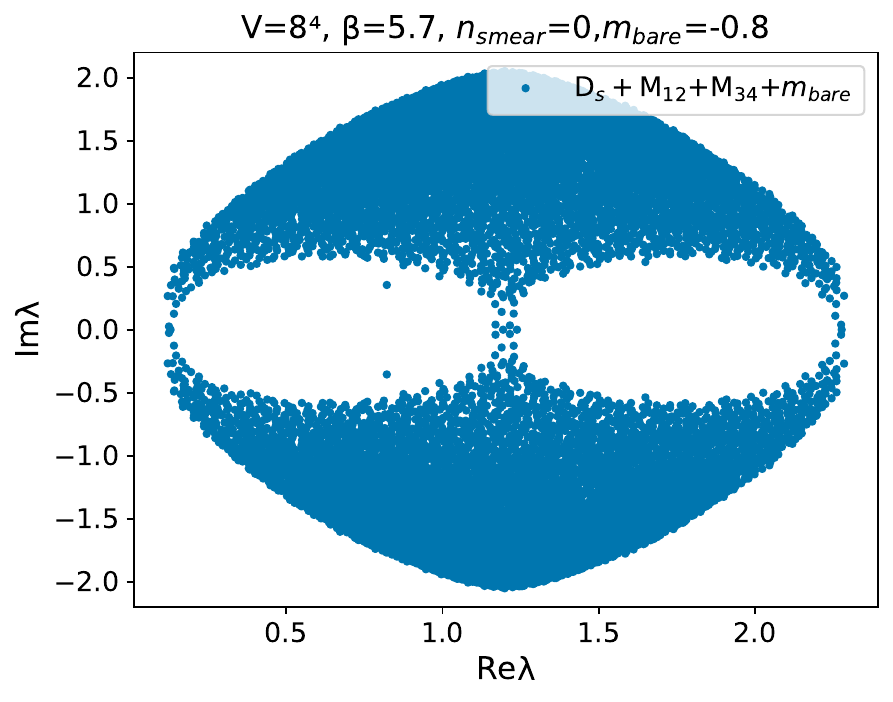}
    \caption{Eigenvalue spectrum of $D_s+(2+M_{12}+M_{34})+m_{bare}$ for $V=8^4$, $\beta=5.7$ and $n_{\textbf{smear}}=0$.}
    \label{fig:EVspectrum}
\end{figure}

\begin{figure}[htbp!]
    \centering
    \includegraphics[width=0.8\textwidth]{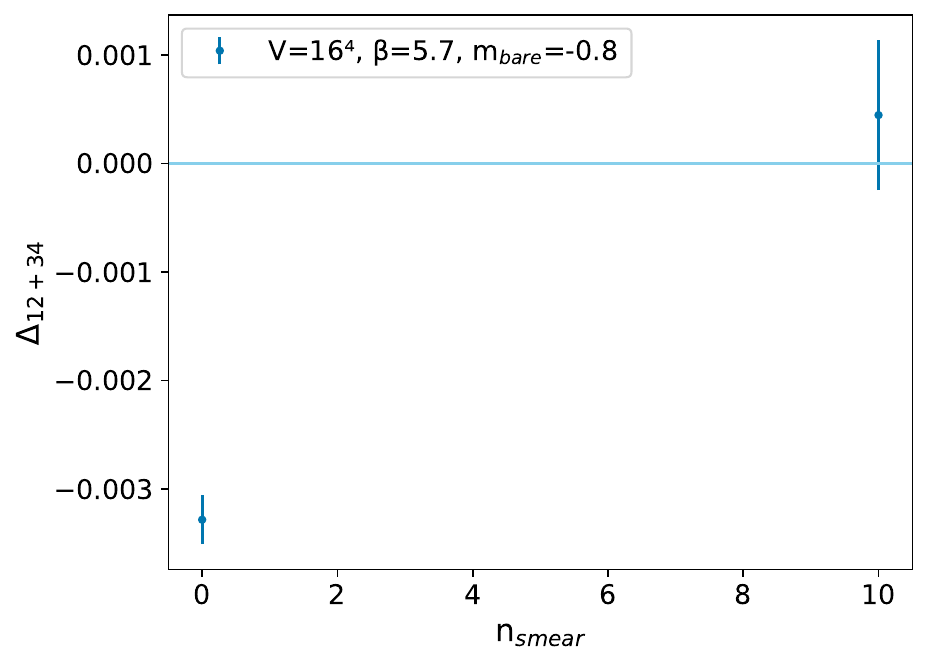}
    \caption{Gluonic counterterm.}
    \label{fig:counterterm}
\end{figure}

\begin{figure}[htbp!]
    \centering
    \includegraphics[width=0.8\textwidth]{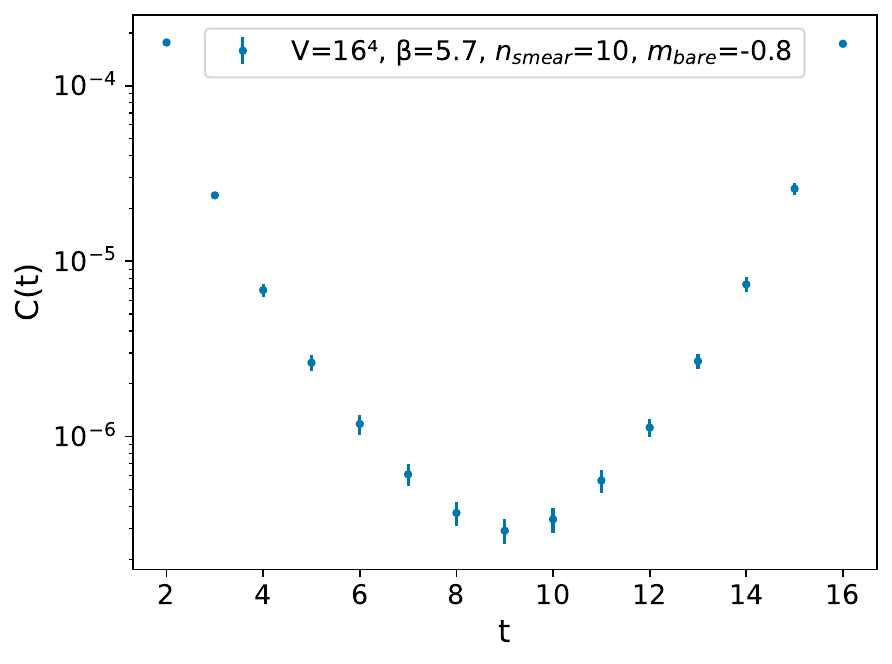}
    \caption{Pseudoscalar propagator.}
    \label{fig:propagator}
\end{figure}

\begin{figure}[htbp!]
    \centering
    \includegraphics[width=0.8\textwidth]{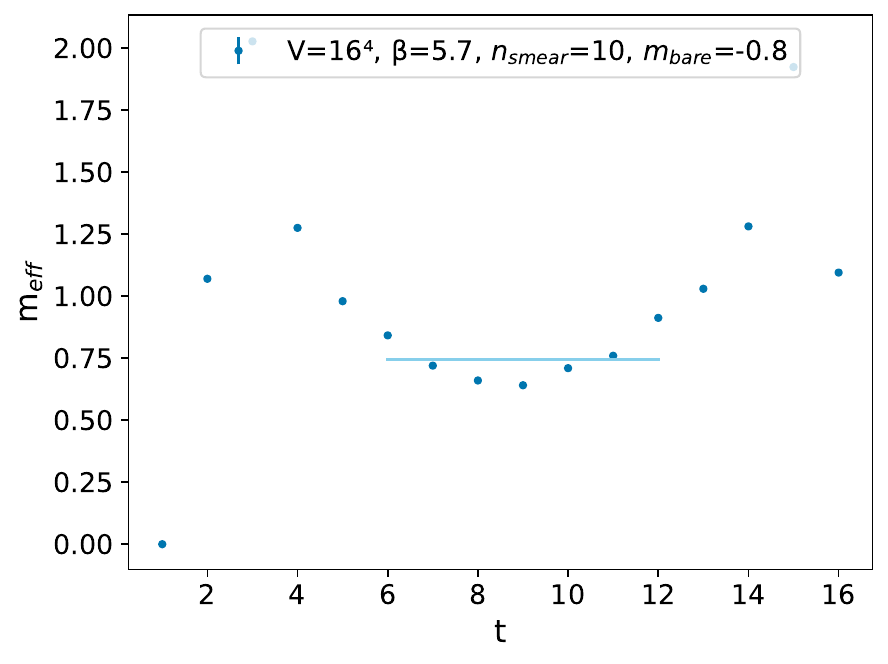}
    \caption{Effective mass.}
    \label{fig:effectivemass}
\end{figure}

The results presented in this section are preliminary with a small statistic of only up to 10 configurations. An exemplary eigenvalue sprectrum for an $8^4$ lattice is given in \autoref{fig:EVspectrum}. The rotational symmetry breaking is visible at the two branches that the doublers split into where eigenvalues are not symmetric with respect to the middle branch. 
\\
Measuring the counterterms that appear from rotational symmetry breaking in the dynamical simulations, we check the averaged difference
\begin{align}
    \Delta_{12+34}=\frac{1}{2}\left(U_{12}+U_{34}\right)-\frac{1}{4}\left(U_{13}+U_{23}+U_{14}+U_{24}\right)
    \label{eq:counterterm_measured}
\end{align}
as shown in \autoref{fig:counterterm}. The counterterms are negligible, just as was observed for pure gauge configurations \cite{PoS2024}.
\\
In addition to the counterterms we compute a physical observable which is the pseudoscalar propagator 
\begin{align}
        C(t) = \langle O(n)\Bar{O}(m_0)\rangle =\sum_{n}\abs{D_{sW}^{-1}(n,m_0)}^2
    \end{align}
where  $O(n) = \Bar{u}(n)\epsilon d(n)$ and $D_{sW} = D_{H}+2+m_{bare}$ as shown in \autoref{fig:propagator} and the corresponding mass from the plateau-fit in \autoref{fig:effectivemass}. 
\\
The mass is fitted in the range $t\in[6,12]$ resulting in $am_{\text{eff}}=0.7433(21)$. For the lattice spacing of $a=0.0843(29)\SI{}{\femto\meter}$ from \autoref{fig:scalesetting}, the mass is determined to be $m=\SI{1740(59)}{\mega \electronvolt}$.

\section{Conclusion}
We have studied the applicability of staggered fermions with a taste splitting mass term, which is the single taste operator, in dynamical simulations. So far it is safe to say that the splitting of the tastes are as expected and confirmed by the eigenvalues spectrum. The gluonic counterterms are still insignificant numerically as was observed in pure gauge SU(3) simulations \cite{PoS2024}. Physical quantities, such as the pion propagator behave as expected, nevertheless they need to be investigated and tunned further as the extracted pion mass is very high.   

\section*{Acknowledgement}
We thank Stephan Durr, Timo Eichhorn and Tatsuhiro Misumi for insightful discussions. Nuha Georgiev-Chreim is supported by the Hans-Böckler-Stiftung under grant No. 415369. This work is supported by the DFG,grant No. HO 4177/1-1.

\end{document}